\documentclass[11 pt]{iopart} 

\usepackage[numbers,sort&compress]{natbib}
\usepackage{amsfonts} 
\usepackage{amssymb} 
\usepackage{graphicx}
\usepackage[british]{babel} 

\def\newblock{\hskip .11em plus .33em minus .07em} 
\def\GWLC{{\sc Gwlc}} 
\def\WLC{{\sc Wlc}} 
\def\eps{\mathcal{E}}

\begin{document} 

\title{Inelastic mechanics of sticky biopolymer networks}

\author{Lars Wolff$^1$, Pablo Fernandez$^2$, and Klaus Kroy$^1$} 

\address{$^1$Institute
for Theoretical Physics, University of Leipzig, Vor dem Hospitaltore 1,
04102 Leipzig, Germany} 

\address{$^2$Department of Physics, TU M\"unchen, James-Frack-Str.,
85748 Garching, Germany}

\ead{lars.wolff@itp.uni-leipzig.de}

\begin{abstract} 

	We propose a physical model for the nonlinear inelastic mechanics of
	sticky biopolymer networks with potential applications to inelastic cell
	mechanics. It consists in a minimal extension of the glassy wormlike chain
	(\GWLC) model, which has recently been highly successful as a quantitative
	mathematical description of the viscoelastic properties of biopolymer
	networks and cells. To extend its scope to nonequilibrium situations,
	where the thermodynamic state variables may evolve dynamically, the \GWLC\
	is furnished with an explicit representation of the kinetics of breaking and
	reforming sticky bonds.  In spite of its simplicity the model exhibits
	many experimentally established non-trivial features such as power-law
	rheology, stress stiffening, fluidization, and cyclic softening effects.

\end{abstract}


\submitto{New Journal of Physics} 

\section{Introduction} 

\label{sec:introduction} 

In many studies of cell mechanics and dynamics, the cell is
characterized as a viscoelastic body
\cite{Fung:1993,MacKintoshKaes:1995,KruseJoanny:2004}.  It is an
intriguing question to what extent such mechanical behaviour can be
rationalized in terms of comparatively simple polymer physics
models. In this respect, the comparison of cell rheological data and
minimalistic \emph{in vitro} reconstituted constructs of the
cytoskeleton, such as pure actin solutions
\cite{GislerWeitz:1999,MasonGisler:2000,GardelValentin:2003,SemmrichStorz:2007,SemmrichLarsen:2008}
or crosslinked actin networks
\cite{GardelNakamura:2006,TharmannClaessen:2007,LuanLieleg:2008,ClaessensSemmrich:2008,KaszaNakamura:2009,KoenderinkDogic:2009},
has recently provided many new insights.  Important progress has also
been achieved in the development of phenomenological mathematical
descriptions.  This includes approaches related to the tube model
\cite{KasStrey:1994,Morse:1999,Morse:2001,HinschWilhelm:2007},
tensegrity-based approaches
\cite{CanadasLaurent:2002,SultanStamenov:2004,CanadasWendling:2006},
effective-medium models
\cite{GittesMacKinto:1998,StormPastore:2005,BroederszStorm:2008}, and
some others \cite{AstromKumar:2008,KollmannsbergerFabry:2009}.
In particular, the glassy wormlike chain (\GWLC) model
\cite{KroyGlaser:2007}, a phenomenological extension of the standard
\WLC\ model of semiflexible polymers \cite{Kroy:2008} has been
successful in describing a plethora of rheological data for polymer
solutions \cite{SemmrichStorz:2007,GlaserHallatsc:2008} and living
cells \cite{KroyGlaser:2009} over many decades in time with a minimum
of parameters. However, all these studies were primarily concerned
with \emph{viscoelastic} behaviour, while the latest investigations
have underscored the glassy
\cite{FabryFredberg:2003,DengTrepat:2006,ZhouTrepat:2009} fragile
\cite{TrepatDeng:2007,KrishnanPark:2009}, and inelastic
\cite{TrepatDeng:2007,KrishnanPark:2009,FernandezOtt:2008} character
of the mechanical response of living cells.  Even for biopolymer
networks \emph{in vitro}, experiments operating in the nonlinear
regime had so far to resort to special protocols that minimize plastic
flow \cite{XuTseng:2000,SemmrichLarsen:2008} in order to make contact
with dedicated theoretical models.

The aim of the present contribution is to overcome this restriction by
extending the \GWLC\ to situations involving inelastic deformations. As a
first step, we concentrate onto  \emph{reversible inelastic} behaviour,
where the deformation does not alter the microscopic ground state.   The
protocol applied by Trepat \emph{et al.}~\cite{TrepatDeng:2007} provides a
paradigmatic example. Cells are subjected to a transient stretch such
that, after some additional waiting time in the unstretched state, the
(linear) material properties of the initial state are recovered. The
simplification for the theoretical modelling results from the assumption
that not only the macro-state but also the micro-state of the system may
to a good approximation be treated as reversible under such conditions;
i.e., we assume that the complete conformation of the polymer network,
including the transiently broken bonds between adjacent polymers, is
constrained to eventually return to its original equilibrium state. For
the time-delayed hysteretic response of the network to such protocols one
could thus still speak of a viscoelastic (``anelastic'') response in an
operational sense, but we refrain from doing so in view of the
fundamentally inelastic nature of the underlying stochastic process --- in
contrast to the reversible softening effects observed in
\cite{ChaudhuriParekh:2007}, for example. Indeed, by simply allowing bonds
to reform in new conformational states, the model developed below can
readily be extended to arbitrary irreversible plastic deformations, as
will be demonstrated elsewhere \cite{wolff-bullerjahn-kroy:tbp}.  Before
entering the discussion of our model, we would also like to point out that
the proposed (inelastic) extension of the \GWLC\ is strongly constrained
by identifying the newly introduced parameters with those of the
original (viscoelastic) \GWLC\ model, where possible. Despite its increased
complexity, the extended model will therefore enable us to subject the
underlying physical picture to a more stringent test than hitherto
possible by comparing its predictions to dedicated experiments.  Moreover,
unlike current state-of-the-art simulation studies \cite{AstromKumar:2008}
it is not limited to rod networks but is firmly routed in a faithful
mathematical description of the underlying Brownian polymer dynamics.

This paper is organized as follows. First, we review some basic facts
about the \GWLC\ in section \ref{sec:gwlc}. Next, in section
\ref{sec:interaction}, we introduce our extended reversible
  inelastic version, which we formulate using the notion of an
effective interaction potential as in the original construction
of the \GWLC\ in \cite{KroyGlaser:2007}.  (A preliminary account of
the basic procedure and some of its cell-biological motivation
including reversible bond-breaking kinetics has recently been given in
a conference proceedings \cite{WolffKroy:2009}.)  Sections
\ref{sec:viscoelastic} and \ref{sec:fluidization} explain the physical
mechanism underlying the mechanical response under pulsed and
periodically pulsed loading, while section \ref{sec:remodelling}
illustrates its phenomenology. We demonstrate that the model exhibits
the hallmarks of nonlinear cell mechanics: strain/stress stiffening,
fluidization, and cyclic softening
\cite{TrepatDeng:2007,KrishnanPark:2009,FernandezOtt:2008}.  Section
\ref{sec:intr_lengths} investigates the relevance of the lately
quantified structural heterogeneities in networks of semiflexible
polymers \cite{GlaserChakrabo:2009} for the mechanical properties,
before we conclude and close with a brief outlook.

\section{Theory} 

\subsection{The glassy wormlike chain}

 \label{sec:gwlc}

The glassy wormlike chain (\GWLC) is a phenomenological extension of the
wormlike chain (\WLC) model, the well-established standard model of
semiflexible polymers. A broad overview over \WLC\ and \GWLC\ dynamics can
be found elsewhere \cite{Kroy:2008}. The \WLC\ describes the mechanics of
an isolated semiflexible polymer in an isothermal viscous solvent. In the
weakly bending rod approximation, a solution of the stochastic
differential equations of motion for the \WLC\ is possible {\it via} a
mode decomposition ansatz for the transverse displacement of the polymer
contour from the straight ground state. The individual modes labelled by
an index $n$ are independent of each other and decay exponentially with
rates $\tau_n^{-1}$. For convenience, we set the thermal energy $k_B T=1$,
so that the bending rigidity can be identified with the persistence length
$\ell_p$, in the following. Using this convention, the \WLC\ expression
for the transverse susceptibility of a polymer of contour length $L$ (with
respect to a point force) reads \cite{KroyGlaser:2007}
\begin{equation}
  \alpha_{\mbox{\scriptsize \WLC}}(\omega)= \frac{L^3}{\ell_p \pi^4}
\sum_{n=1}^{\infty} \frac1{(n^4+n^2 f/f_L )(1+i \omega \tau_n)}.
\label{eq:sus_wlc}
\end{equation}
Here, $f$ is an optional backbone tension, and $f_L\equiv
\pi^2\ell_p/L^2$ the Euler buckling force for a hinged rod of length
$L$. The different powers of $n$ in the denominator give notice
  of the competition of bending and stretching forces.

\begin{figure}[h] \begin{center}
	\includegraphics[width= 9 cm]{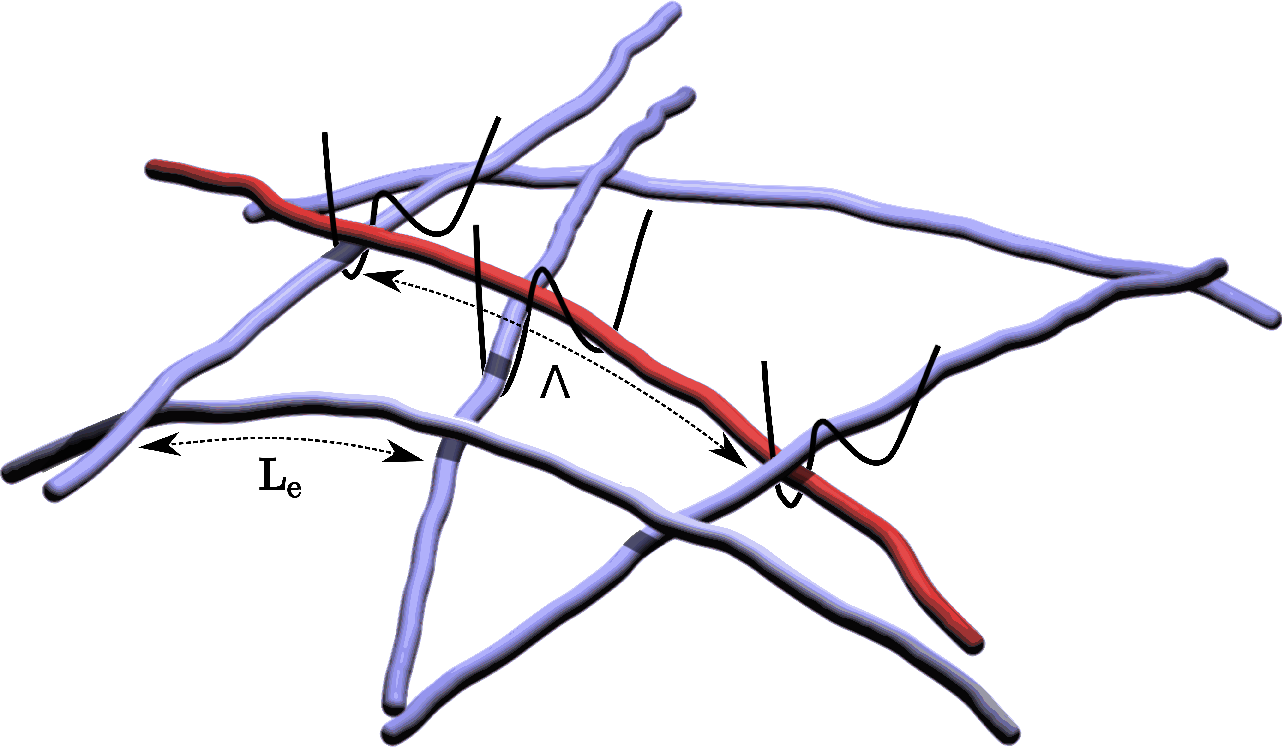}
        \caption{Cartoon of a test polymer (differently coloured)
            in a network. The trapping of the test polymer is
            schematically represented by means of an effective
            interaction (figure~\ref{fig:potential}) indicated by the
            potential wells at sticky entanglement points, which are on
            average separated by an contour length $L_e$.  The test polymer
            can bind/unbind by overcoming an energy barrier of height
            $\eps$.  The average contour length between the closed
            bonds is $\Lambda$.}
	\label{fig:cartoon}
\end{center}\end{figure}

In the \GWLC, interactions of the polymer with the surrounding
network (e.g.\ excluded volume interactions or sticky contacts) are
reflected in an altered \WLC\ mode relaxation spectrum
\cite{KroyGlaser:2007,GlaserHallatsc:2008}.  The intuitive --- albeit not
fully microscopic --- picture underlying the formulation of the \GWLC\ is
illustrated in figure~\ref{fig:cartoon}, which depicts a test polymer
reversibly bound to the background network \emph{via} potential wells at
all topological contacts, the so-called entanglement points. Consider now
a generic point somewhere along the polymer backbone (not an entanglement
point). It can relax freely until the constrictions are being felt, which
slow down the contributions from long wavelength bending modes.  The
\GWLC\ translates this intuition into a simple prescription for the mode
spectrum: the short-wavelength modes are directly taken over from the
wormlike chain model, while modes of wavelength $\lambda_n$ longer than
the typical contour length $\Lambda$ between adjacent bonds are modified to
account for the slowdown. Motivated by the physical picture illustrated in
figure \ref{fig:cartoon}, the slowdown of the relaxation of a wavelength
$\lambda_n$ is expressed by an Arrhenius factor $\exp\left[\eps\left(
\lambda_n/\Lambda-1\right)\right]$ for breaking $(\lambda_n/\Lambda-1)$
potential energy barriers of height $\eps$ simultaneously.
Accordingly, the phenomenological recipe to turn a \WLC\ into a \GWLC\
reads:
\begin{equation}
  \tau_n \to 	\tilde{\tau}_n=\left\{ \begin{array}{c c} \tau_n &
\lambda_n < \Lambda \\ 
	\tau_n \exp\left[ \eps(\lambda_n/\Lambda -1)\right] & \lambda_n
\geq \Lambda \end{array}\right. .
\end{equation}
Upon inserting this into (\ref{eq:sus_wlc}) one obtains the \GWLC\
susceptibility $\alpha_{\mbox{\scriptsize \GWLC}}(\omega)$. The
microscopic ``modulus'' for transverse point excitations of a generic
backbone element on a test polymer is then defined as the inverse
susceptibility, $g_{\mbox{\scriptsize
\GWLC}}(\omega)=1/\alpha_{\mbox{\scriptsize \GWLC}}(\omega)$.  An
approximate expression for the macroscopic shear modulus is obtained along
similar lines \cite{GittesMacKinto:1998,KroyGlaser:2007}.

In the original equilibrium \GWLC\ theory, $\Lambda$ was assumed to be
a constant on the order of the entanglement length of the network,
$\Lambda\gtrsim L_e$. Note, however, that $L_e$ is a geometric
quantity (which is determined by the polymer concentration and the
persistence length) while the contour length $\Lambda$ between closed bonds
clearly depends on the state of the bond network. One therefore has to
allow for an increase of $\Lambda$ with the number of broken bonds in
non-equilibrium applications. This issue is explored in the following.

\subsection{Effective interaction potential and bond kinetics}
\label{sec:interaction} 
\begin{figure}[h] \begin{center}
	\includegraphics[width=9 cm]{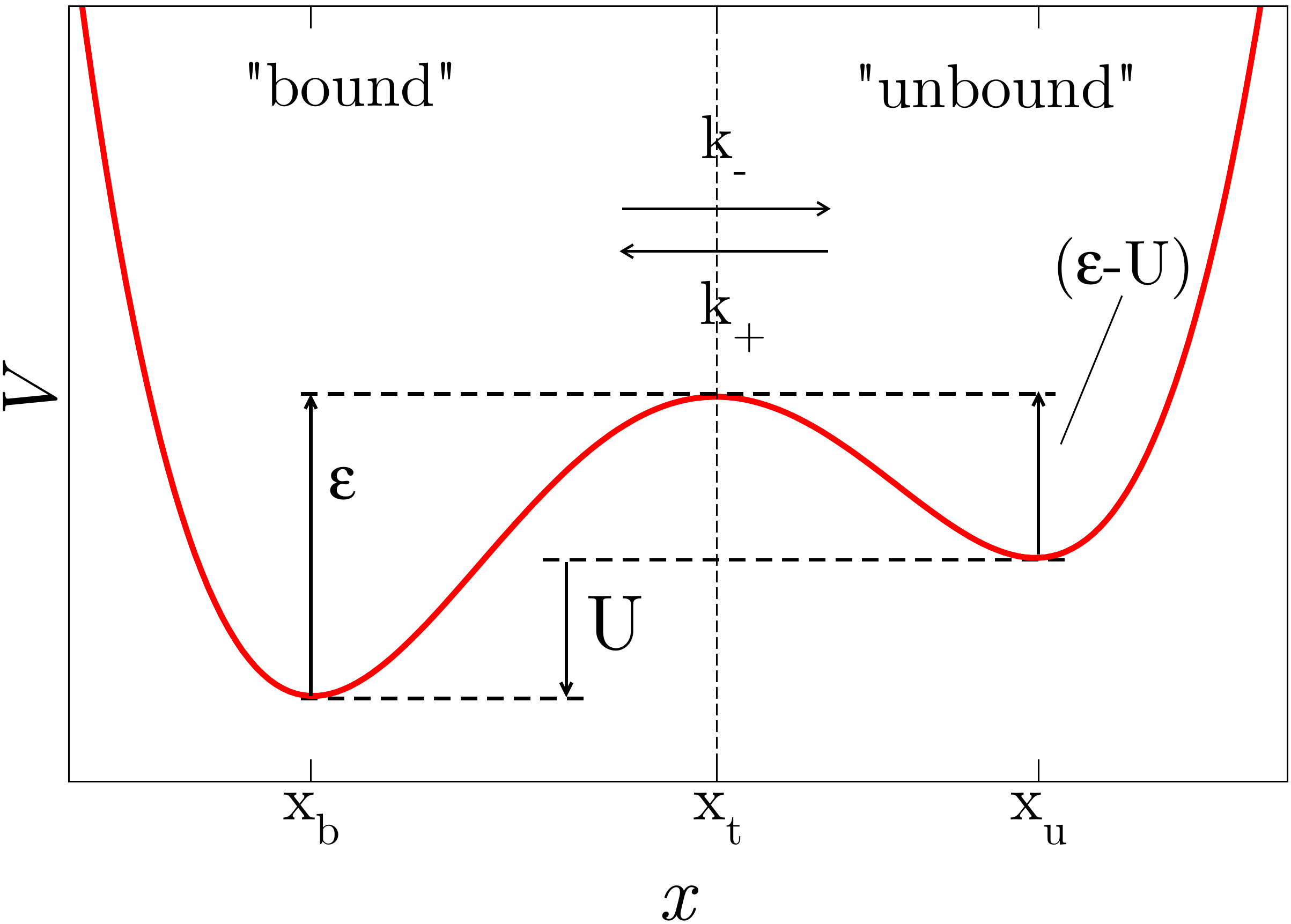}
	\caption{Sketch of the schematic effective interaction
          potential for a test segment against the background.  A
          polymer segment can either be ``bound'' by a transient
          crosslinker or a sticky patch on the backbone of another
          polymer (left potential well) or ``unbound'' and merely
          confined by the surrounding network to a tube like cage
          (right well). When considering an ensemble of contacts, the
          fraction of closed bonds depends on the free energy
          difference $U$ between both states. The transition rates
          $k_-$ and $k_+$ between the bound and unbound state depend
          on the barrier heights $\eps$ and $(\eps-U)$, respectively.
          An externally applied force can tilt the potential and
          favour binding or unbinding events.}
	\label{fig:potential}
\end{center}\end{figure}

All mechanical quantities calculated within the \GWLC\ model crucially
depend on the interaction length $\Lambda$. In previous applications
of the model \cite{SemmrichStorz:2007,KroyGlaser:2007,KroyGlaser:2009}
it was assumed that $\Lambda$ remains constant --- equal to its
equilibrium value and unaffected by the deformation of the sample. In
other words, the equilibrium theory allowed for statistical bond
fluctuations but not for a dynamical evolution of the parameters
characterizing the thermodynamic state of the bond network.  An
obvious starting point for generalizations of the model to
non-equilibrium situations is therefore to consider the number of
closed bonds, and therefore also $\Lambda$, as {\em dynamic
  variables}, dependent on the strain- and stress-history of the
network. We now provide a mean-field description to account for such a
dynamically evolving bond network. For clarity, we return to the
intuitive picture underlying the \GWLC\, where the (possibly
crosslinker- or molecular-motor-mediated) complex interactions between
the polymers are summarized into an effective interaction potential
for a test segment against the background, as sketched in figure
\ref{fig:potential}.  The same idea has also been used before in many
related situations (e.g.\
\cite{Bell:1978,SchwarzErdmann:2006,KollmannsbergerFabry:2009}). 
The generic potential exhibits three essential features: a
well-defined bound state, a well-defined unbound state, and a barrier
in between (figure \ref{fig:potential}). The confining well
corresponding to the unbound state represents the tube-like
caging of the test polymer within the surrounding network
\cite{KasStrey:1994,DichtlSackmann:2002,Morse:2001}.  For ease of
notation, we further introduce the dimensionless fraction of closed
bonds or, {\em bond fraction}
\begin{equation}
  \nu(t) \equiv \Lambda_0/ \Lambda(t) \;,
	\label{eq:nu_lambda}
\end{equation}
which is simply the ensemble-averaged fraction of sticky contacts that
are actually in the bound state. The minimum bond distance $\Lambda_0$
is typically on the order of $L_e$, but may be somewhat larger in
situations where the bonds are mediated by crosslinker molecules or by
partially sterically inaccessible sticky patches (as e.g.\ for helical
molecules \cite{ClaessensSemmrich:2008,GrasonBruinsma:2008}).

A quantitatively fully consistent way of calculating the dynamics of $\nu$
would involve solving the full Fokker-Planck equation for a \WLC-\WLC\
contact --- a formidable program to be pursued elsewhere
\cite{wolff-bullerjahn-kroy:tbp}. Here, for the sake of our qualitative
purposes, we chose to concentrate onto the physical essence of the
discussion, and keep the mathematical structure as simple and transparent
as possible. We therefore approximate the dynamics by a simple exponential
relaxation as familiar from the standard example of reacting Brownian
particles, conventionally described by Kramers theory with a
Bell-like force dependence \cite{EvansRitchie:1997,FenebergWestphal:2001}.
Using this simplification and assuming a schematic interaction potential
as depicted in figure~\ref{fig:potential}, the value of $\nu$ is
determined by a competition of bond breaking and bond formation with
reaction rates $k_-$ and $k_+$, respectively. Both rates are represented
in the usual adiabatic approximation (meaning that the equilibration
{\em inside} the wells is much faster than the barrier crossing and
external perturbations) by \cite{EvansRitchie:1997}
\begin{equation}
 k_-\tau_0=  e^{-\mathcal{E}+(x_t-x_b)f }\;, \qquad k_+ \tau_0=
e^{-\mathcal{E}+U-(x_u-x_t)f},
	\label{eq:kpm}
\end{equation}
where $\tau_0$ is an intrinsic characteristic Brownian time scale for bond
breaking and formation\footnote{Strictly speaking the fluctuations in the
potential wells at $x_b$ and $x_u$ are characterized by different Brownian
time scales depending on the width of the potential wells. At the present
stage, we do not bother to distinguish these time scales nor to fine-tune
their numerical values, and identify them, for the sake of simplicity, with
the entanglement time scale $\tau_{L_e}$ in our numerical calculations.},
and $f$ is the force pulling on the bond. Noting that the fraction of open
bonds is $1-\nu$, we can then write down the following rate equation for
the fraction of closed bonds
\begin{equation}
  \dot{\nu}(t) \tau_0 e^{\mathcal{E}}= -  ( e^{-(x_u-x_t)
f(t)+U}+e^{(x_t-x_b) f(t)}) \nu(t) + e^{-(x_u-x_t)f(t)+U}.
\label{eq:rate}
\end{equation}
 The time dependence of $\nu(t)$ leads via (\ref{eq:nu_lambda}) to an
implicit time dependence of the \GWLC-parameter $\Lambda(t)$ and thereby
of all observables derived from the \GWLC. The time-dependent force $f(t)$
in (\ref{eq:rate}) may be thought to result from an externally imposed
stress protocol or from internal dynamical elements such as molecular
motors setting the network under dynamic stress. Hence, via an
appropriately chosen set of slowly changing state parameters $f(t)$,
$U(t)$, $\eps(t)$, \dots\ the model can in principle accommodate for the
active biological remodelling in living cells and tissues
\cite{WolffKroy:2009}. (For a discussion of the relation between the
microscopic $f$ and the macroscopic stress, see
\cite{GlaserHallatsc:2008}.) Note that for constant force, the stationary
value of $\nu$,
\begin{equation}
	\nu_{\rm stat}=\left(1+\exp\left[-U+(x_u+x_b) f_{\rm
stat}\right]\right)^{-1},
\end{equation}
 obtained by setting $\dot{\nu}\equiv0$, does not depend on
$\eps$, as it should be (the steady state is independent of the transition
state).

\section{Results and discussions} 
\label{sec:results} 

\subsection{Viscoelastic properties}
\label{sec:viscoelastic}

\begin{figure}[h] \begin{center}
	\includegraphics[width=7 cm]{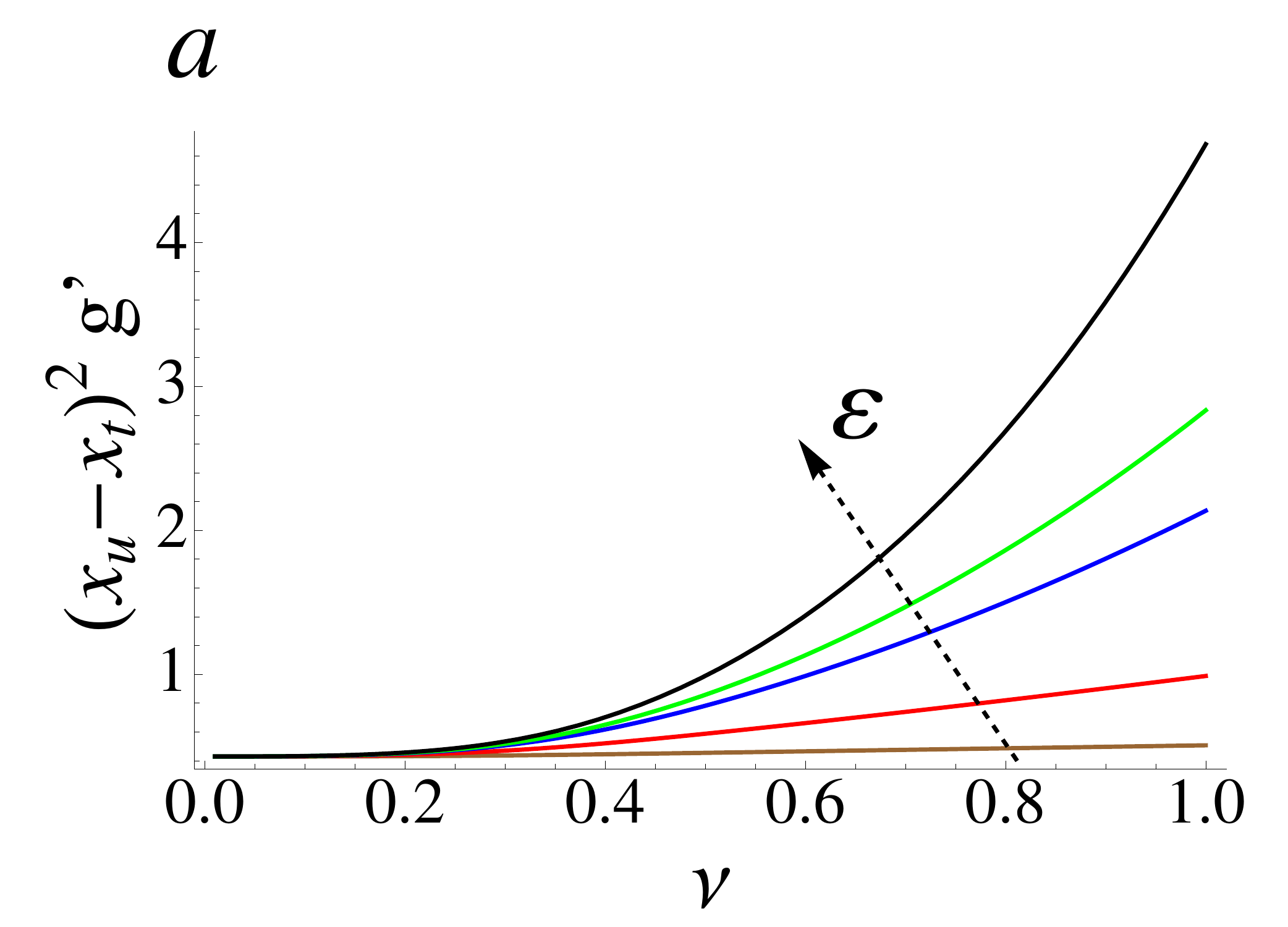}
	\includegraphics[width=7 cm]{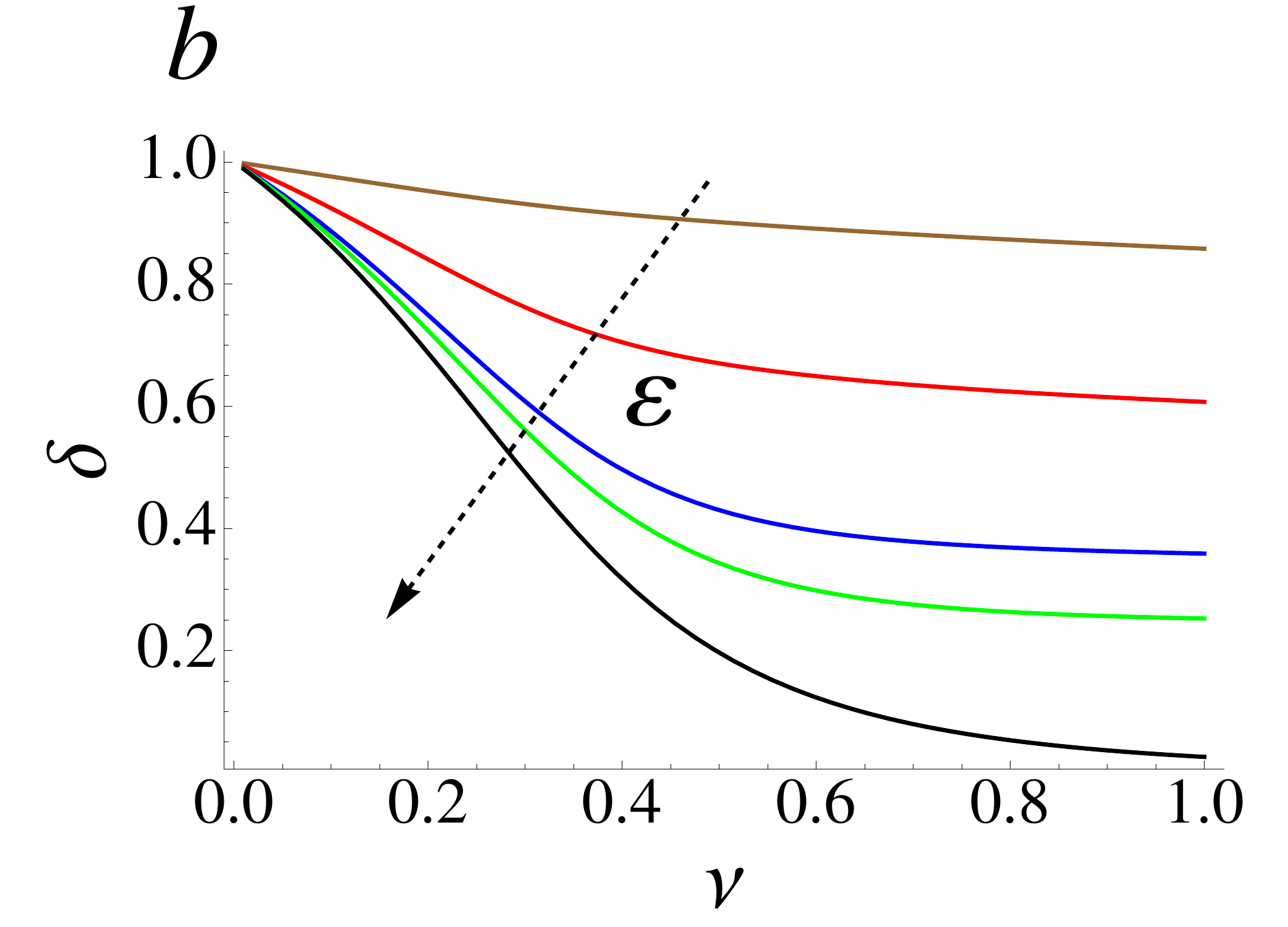}
        \caption{Dependence of the microrheological ``modulus''
          $g_{\mbox{\scriptsize \GWLC}}$ on the bond fraction $\nu$
          for various values of $\eps$ (brown: $\eps=0.1$, red:
          $\eps=1$, blue: $\eps=5$, green: $\eps=10$, black: $\eps=\infty$) at a frequency
$\omega_0 \approx 1/10 \tau^{-1}_{\Lambda_0}$ and a force
$f=f_{\Lambda_0}/2$ ($\tau_{\Lambda_0}$ is the relaxation time of a mode
of length $\Lambda_0$ and $f_{\Lambda_0}$ is the Euler buckling
force for a rod of length $\Lambda_0$); \emph{left:}
          normalized real part $g'$; \emph{right:} loss angle.}
\label{fig:mod_lam}
\end{center}\end{figure}

Whenever the bond kinetics can be disregarded ($\dot\nu=0$), the
viscoelastic properties are simply those of the bare \GWLC\
\cite{KroyGlaser:2007,SemmrichStorz:2007,GlaserHallatsc:2008,KroyGlaser:2009},
which can basically be characterized as \WLC\ short time dynamics
followed by a slow, highly stretched logarithmic relaxation resembling
power-law rheology with a non-universal exponent within the typical
experimental time windows. Strong perturbations (stress or strain)
then give rise to a pronounced stiffening due to the steeply nonlinear
response of semiflexible polymers under elongation
\cite{MacKintoshKaes:1995}. This well established behaviour can be
understood as the canvas against which we aim to discern the
characteristic mechanical signatures of the bond kinetics.

An observable that will be of particular interest in the following is
the complex microscopic ``modulus'' $g\equiv g_{\mbox{\scriptsize
    \GWLC}}(\omega_0,\Lambda_0/\nu,f)$, introduced in the previous
section. It is used to determine the linear as well as (via a
nonlinear update scheme, see \ref{sec:technical}) the nonlinear force
response of the system. To understand how the time-dependence
(\ref{eq:rate}) of $\nu$ affects this important quantity, it is
instructive to first examine the dependence of $g$ on $\nu$. To this
end, we formally consider $\nu$ temporarily as an independent
  variable instead of determining it from (\ref{eq:rate}). Note that
this approach nevertheless makes sense as for a fixed set of
parameters, $\nu$ can still take any value between zero and one. This
is due to the freedom of choosing an initial state, which can be
  imagined to have evolved from the prior deformation history.

For fixed other parameters, both the real part $g'$ (figure
\ref{fig:mod_lam}a) and the imaginary part $g''$ of $g$ increase
monotonically with $\nu$. We emphasize that $g''$ is not simply
proportional to $g'$ and therefore the loss angle
$\delta=\arctan(g''/g')$ also depends on $\nu$ (figure
\ref{fig:mod_lam}b). For small $\nu$, the loss angle is large,
corresponding to fluid-like behaviour. With increasing $\nu$, the
system becomes more solid-like.  Increasing the barrier height $\eps$
makes the dependence of $g$ on $\nu$ more pronounced (figure
\ref{fig:mod_lam}). Conversely, as can be expected, the
  dependence on the barrier height $\eps$ vanishes with decreasing
  bond fraction $\nu$.  Note that due to the boundary conditions 
the limit of a Newtonian fluid ($\delta=\pi/2$) is not recovered when
formally taking the limit of vanishing bond fraction ($\nu\to0$,
$\Lambda\to\infty$). 
Finally, the potential difference $U$ solely determines the dynamics
of $\nu$ via (\ref{eq:rate}), and therefore influences the modulus $g$
solely through the equilibrium value for $\nu$.

After these general considerations, we now concentrate on the
  nonlinear and non-equilibrium dynamic response of the extended
  inelastic \GWLC\ model, which results from the coupled relaxation of
  the viscoelastic polymer network \emph{and} the transient bond
  network.

\subsection{Stress-stiffening versus rate-dependent yielding}

\label{sec:fluidization}

A convenient way to characterize the mechanical properties of an inelastic
material is a force-displacement curve. For a perfectly linear elastic
(Hookean) body, it would simply consist of a straight line, whereas a
perfectly plastic body would feature as a rectangle delineated by a yield
force $f_{\rm y}$ and an arbitrary plastic strain value. For our
qualitative purpose, we identify the average transverse displacement of
the test polymer segment (which is used to determine the force response,
see \ref{sec:technical}) with the reaction coordinate $x$ of the schematic
potential sketched in figure \ref{fig:potential}; and we identify the
force $f$ entering the reaction rates with the backbone tension of the
test polymer.  As a characteristic length scale for the transverse
displacements we use the width $(x_u-x_t)$ of the effective confinement
potential, which is a measure of a typical mean-square displacement of the
polymer in the unbound state. Using this convention, we now consider the
effect of a time-symmetric displacement pulse on the evolution of
the force $f(t)$ and bond fraction $\nu(t)$. For technical convenience, we
use a Gaussian shape for the displacement pulses, but the qualitative
conclusions to be drawn are largely independent of the precise protocol.
The duration of the displacement pulse, which sets the time scale for the
dynamic response, is used as the unit of time in the following. Here, we
are not interested in short-time tension-propagation and single-polymer
dynamics \cite{HallatschekFrey:2007}, hence a lower bound for pertinent
pulse durations is provided by the interaction time scale
$\tau_{\Lambda_0}\simeq \tau_0$. On the other hand, for pulse durations
longer than $\tau_{0} e^\eps$ the system deformation would be quasistatic
so that no genuinely dynamic effects of the bond kinetics could be
observed.

\begin{figure}[h] \begin{center}
	\includegraphics[width=7 cm]{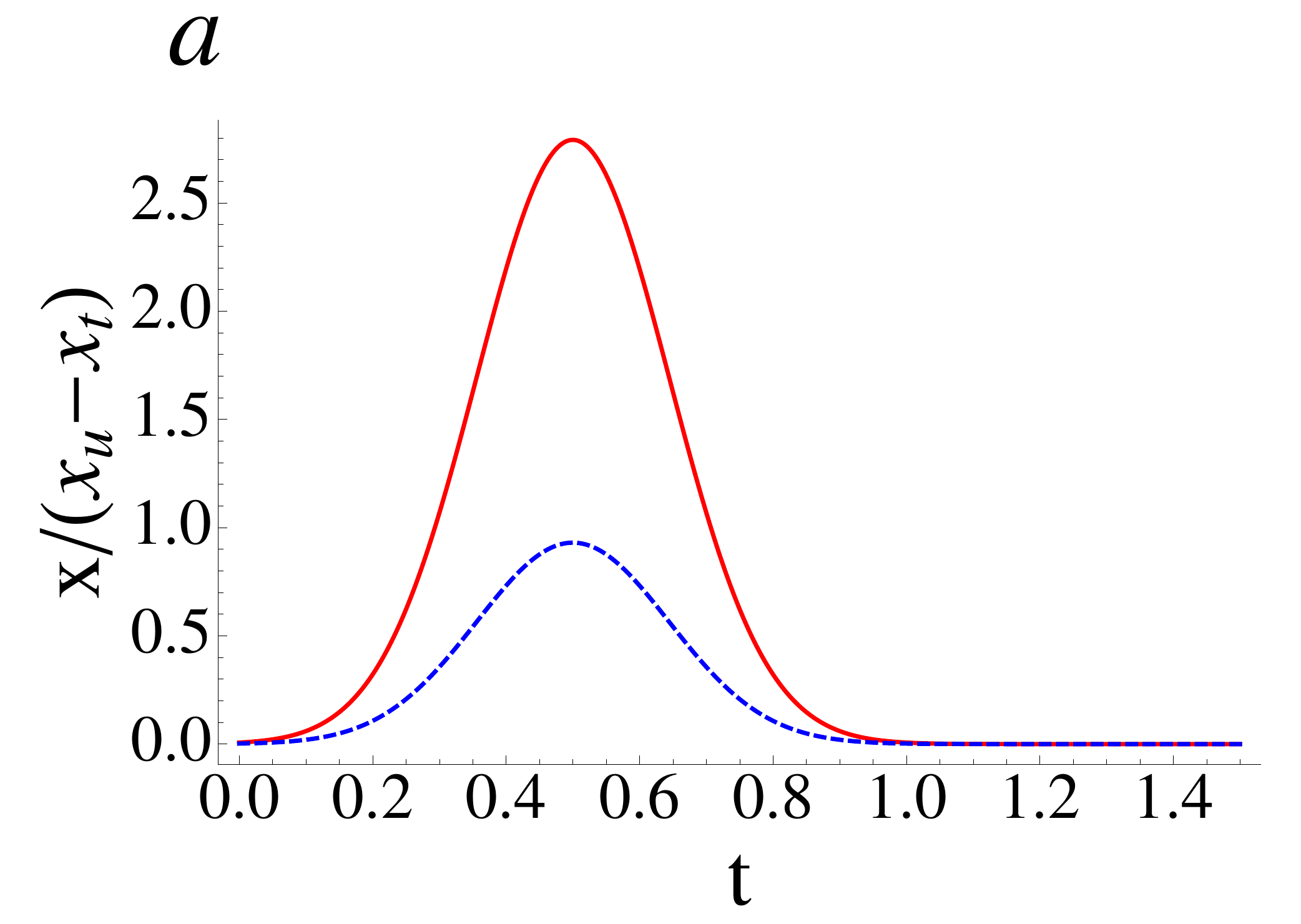}
	\includegraphics[width=7 cm]{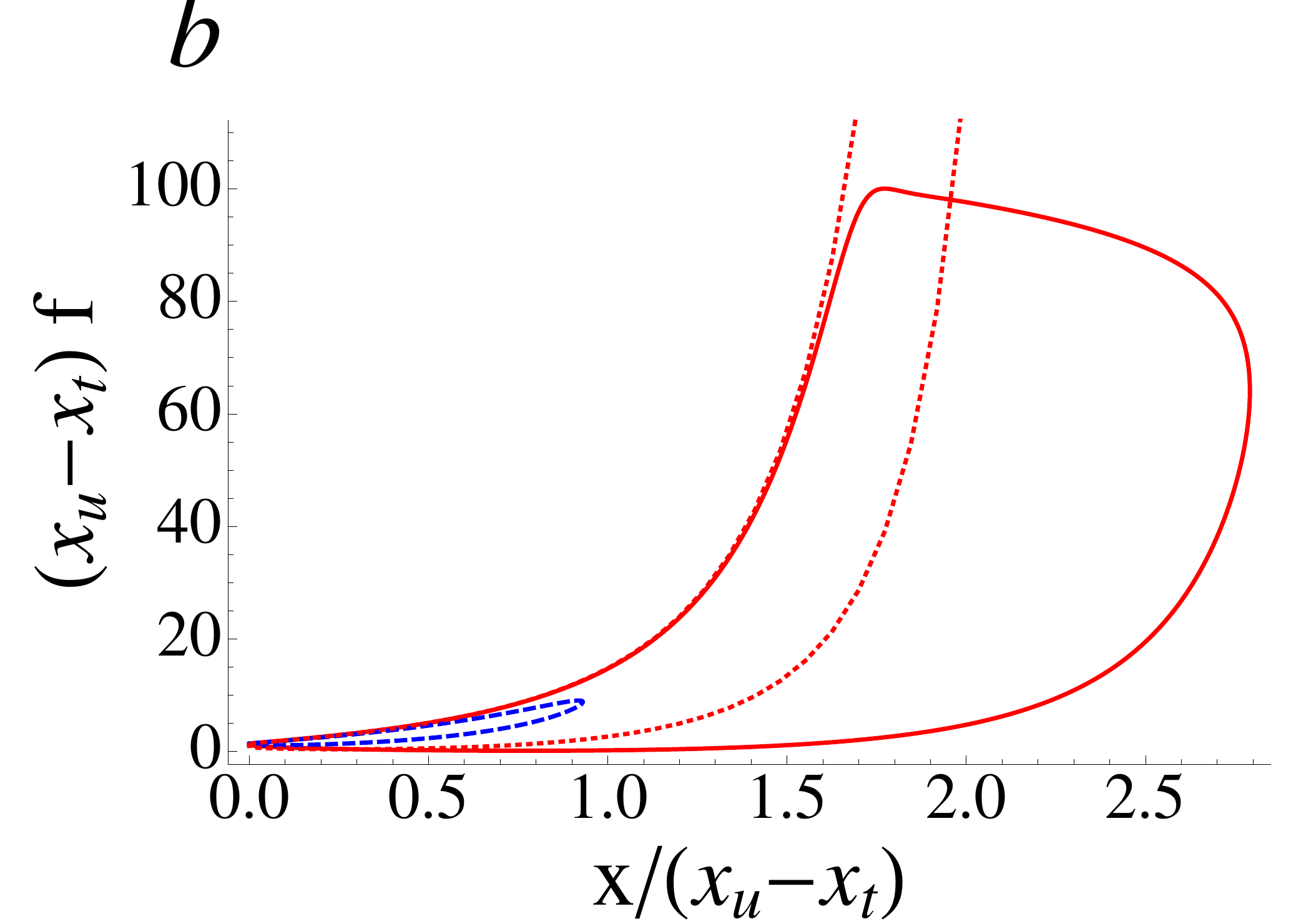}
	\includegraphics[width=7 cm]{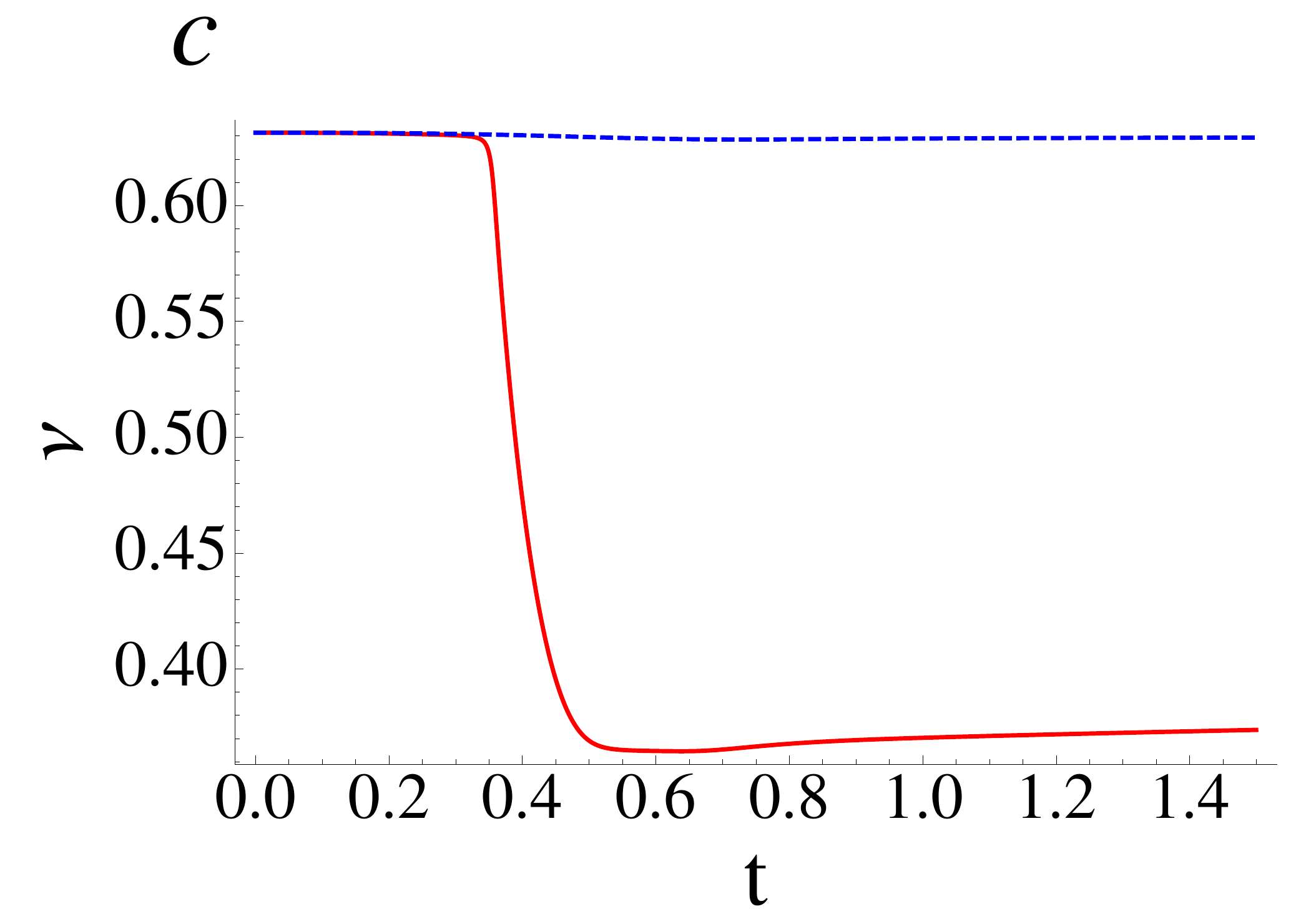}
	\includegraphics[width=7 cm]{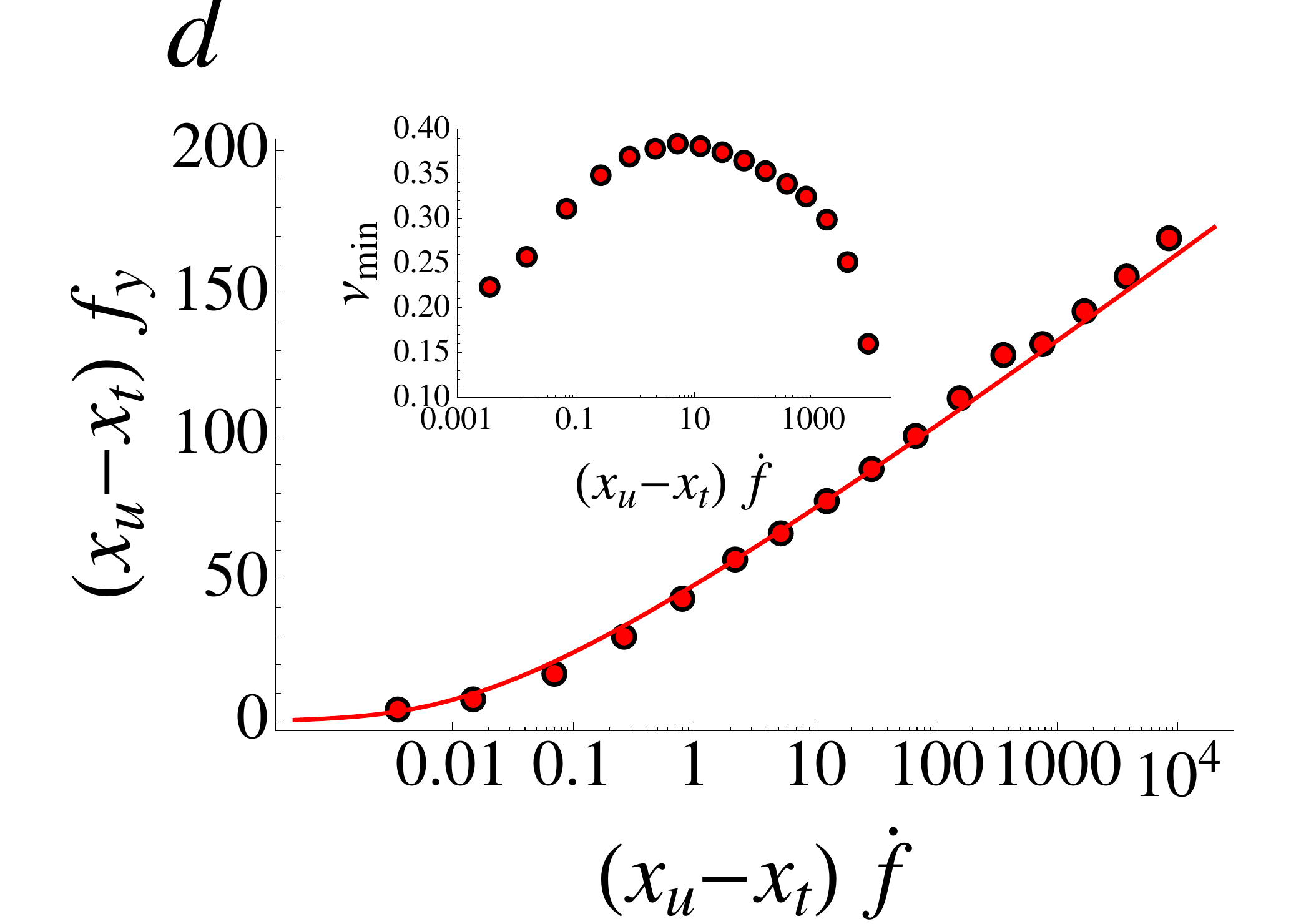}
	\caption{Transition from a stiffening to a softening response
          at a rate-dependent, operationally defined yield force
          $f_{\rm y}$: Gaussian deformation pulses ($a$) with relative
          displacement amplitudes of $0.73$ (dashed) and $2.2$
          (solid); corresponding force-displacement curves and bond
          fraction evolutions ($b$, $c$; dotted curve in $b$: 
response of a \GWLC\ without bond breaking); order-of-magnitude estimate
          for the rate-dependence of the yield force,
          (\ref{eq:rate_fc}) with const.$=2.2$, versus the numerically
          obtained maximum forces of the force-displacement curves for
          various mean force rates ($d$); the minimum bond fraction
          ($d$, inset) depends non-monotonically on the force
          rate; $\eps=10.5$, $U=2$, $\tau_0=0.004$,
          $(x_t-x_b)/(x_u-x_t)= 0.07$, $\Lambda_0=22.3$. }
	\label{fig:force_displacement}
\end{center}\end{figure}

For a small Gaussian displacement pulse of relative amplitude $0.73$
(c.f.\ figure~\ref{fig:force_displacement}a and \ref{sec:technical}), the
shape of the curve predicted by our inelastic \GWLC\ model shows all
features of a {\em viscoelastic} medium (figure
\ref{fig:force_displacement}b, blue dashed curve). It starts with a nearly
linear regime for relative displacements $\lesssim 0.4$, where a weak
stiffening sets in. Due to dissipation in the medium, the path back to the
initial state takes its course at lower force, which causes a weak
hysteresis.  This is essentially the viscoelastic response that already
the bare \GWLC\ model would have predicted.
For a larger relative deformation amplitude of $2.2$, however, the
predictions of the bare \GWLC\ and the inelastic \GWLC\ diverge. The
bare \GWLC\ predicts a strong stiffening (figure
\ref{fig:force_displacement}b, red dotted curve), whereas the full
model exhibits an initial stiffening regime followed by a pronounced
softening (figure \ref{fig:force_displacement}b, solid red curve). A
very interesting observation is that ``softening'' in this case does
not only mean a decrease of the slope of the force-extension curve,
but that the slope actually becomes {\em negative} over an extended
parameter region, once an operationally defined threshold force
$f_{\rm y}$ is exceeded. This ``flow state'' bears more resemblance to
plastic flow than to viscoelastic relaxation. The reason for this
qualitatively new phenomenon is the force-induced bond breaking.
Theoretically this is best exemplified by the time-dependence of the
(experimentally not easily accessible) fraction $\nu(t)$ of closed
bonds (figure \ref{fig:force_displacement}c).  A glance at the bond
fraction during the weak deformation scenario
(figure~\ref{fig:force_displacement}c, blue dashed curve) reveals how
the limit of a bare \GWLC\ is recovered, namely whenever the
deformation is not sufficiently violent and persistent to
significantly decrease the fraction of closed bonds.  For the large
deformation scenario, in contrast, the bond fraction stays only
initially constant (figure~\ref{fig:force_displacement}c, solid red
curve). As a consequence of the large strain the force rises steeply,
as can be seen from the strong stiffening in
figure~\ref{fig:force_displacement}b (solid red curve).  At the yield
force, the bond fraction suddenly decreases to nearly half of its
initial value during a very short time. The decrease in the bond
fraction is accompanied by a somewhat slower drop in the force, which
is reflected in the sudden softening of the force-displacement
curve. The bond fraction eventually recovers on a much slower time
scale, which is roughly given by $\tau_0 \exp(\eps-U)$.  Note that
even though the effects of the inelastic response dominate the
stress-strain curve, the viscoelastic relaxations from the underlying
\GWLC\ model are still present. They could hardly be disentangled
  from the inelastic contributions, though, without an underlying faithful model of the
  viscoelastic response at hand.
%
%

In summary, we observe a competition between force-amplitude dependent
stress-stiffening and force-rate dependent yielding events. If the
backbone force stays much smaller than the yield force, the adaptation
of the bond network requires an adiabatically long time, on the order
of the bond lifetime $\tau_0e^{\eps}$, so that plain viscoelastic
\GWLC\ behaviour is observed on the pulse time scale. In contrast, if
the backbone tension $f$ reaches the yield force $f_{\rm y}$, the bond
fraction declines sharply, thereby switching the response from
the \WLC/\GWLC-typical stiffening to a pronounced softening.

The rate dependence of the yield force $f_{\rm y}$ can be estimated by
setting the time it takes to reach the yield force, $f_{\rm y}/\dot f$,
equal to the force-dependent time scale of bond opening,
$k^{-1}_{-}(f)$, from (\ref{eq:kpm}),
\begin{equation}
  f_{\rm y}\dot{f}^{-1} \simeq  k_{-}^{-1}(f_{\rm y}) \quad \Rightarrow \quad (x_t-x_b)f_{\rm
y}  \approx  {\rm LW}\left(\mbox{const.} \times (x_t-x_b) \tau_0
e^\eps \dot f \right)  .
	\label{eq:rate_fc}
\end{equation}
Here, LW$(x)$ is the positive real branch of the Lambert W function. In
figure \ref{fig:force_displacement}d, this estimate is compared with
results from numerical evaluations for Gaussian displacement pulses at
different average rates. Apart from the numerical errors, the slight
deviations from the estimate may be attributed to the fact that the force
rate is not constant for the Gaussian protocol. They can be mostly
eliminated by using force ramp protocols of various slopes instead of the
Gaussian displacement pulses. For not too low rates, the rate dependence
can be approximated by the even simpler relation
\begin{equation}
	(x_t-x_b)f_{\rm y}  \approx  (\eps +\ln\dot f \tau_0 +
\mbox{const.})  \;,
\end{equation}
where the force rate has to be normalized by a suitable force scale.

The minimum bond fraction reached during the application of the
  pulse, which we interpret as a measure of the degree of
  fluidization, depends {\em non-monotonically} on the rate (figure
  \ref{fig:force_displacement}d, inset). Qualitatively, this can be
  understood by noticing that two different factors influence the
  fluidization, namely the maximum force attained during deformation
  and the time over which the force is applied. The maximum force is
  simply the yield force $f_{\rm y}$, with the rate dependence in
  (\ref{eq:rate_fc}) and figure \ref{fig:force_displacement}d, while
  the reciprocal rate itself sets the time scale. For high rates
  $(x_u-x_t)\dot{f}\gtrsim 1$ (in our example) the rate-dependence of
  the maximum force wins and the minimum $\nu$ reached decreases with
  increasing rate. For low rates, where the rate-dependence of the
  maximum force is much weaker (figure \ref{fig:force_displacement}d)
  the minimum $\nu$ decreases with increasing pulse duration,
  \emph{viz.}\ decreasing rate. Note that for slow pulses (low rates),
  the bond fraction after the pulse may be quite different from the
  minimum $\nu$, as significant recovery may already take place during
  the pulse.

\subsection{Transient remodelling and cyclic softening}

\label{sec:remodelling}

\begin{figure}[h] \begin{center}
	\includegraphics[width=7 cm]{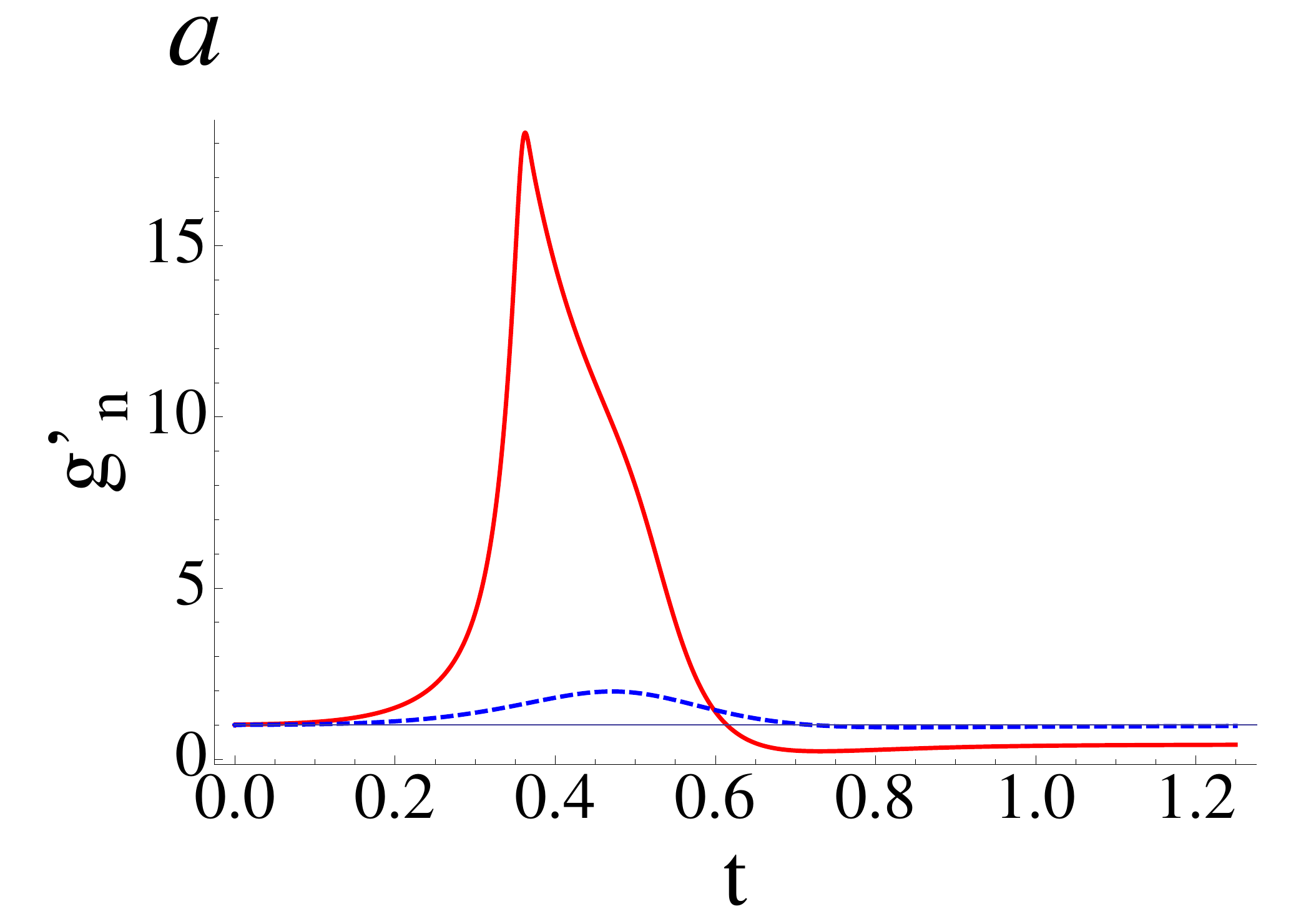}
	\includegraphics[width=7 cm]{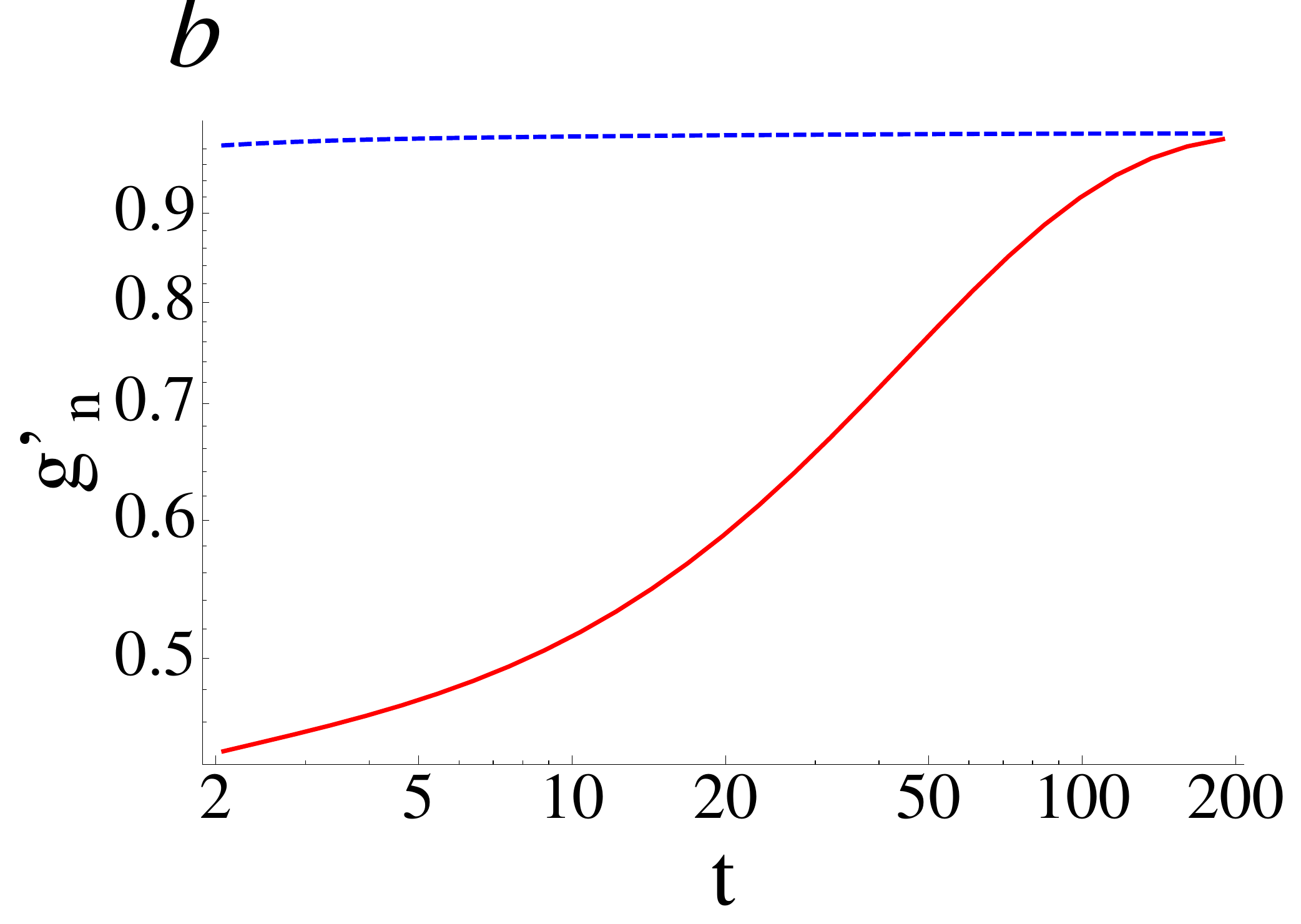}
	\includegraphics[width=7 cm]{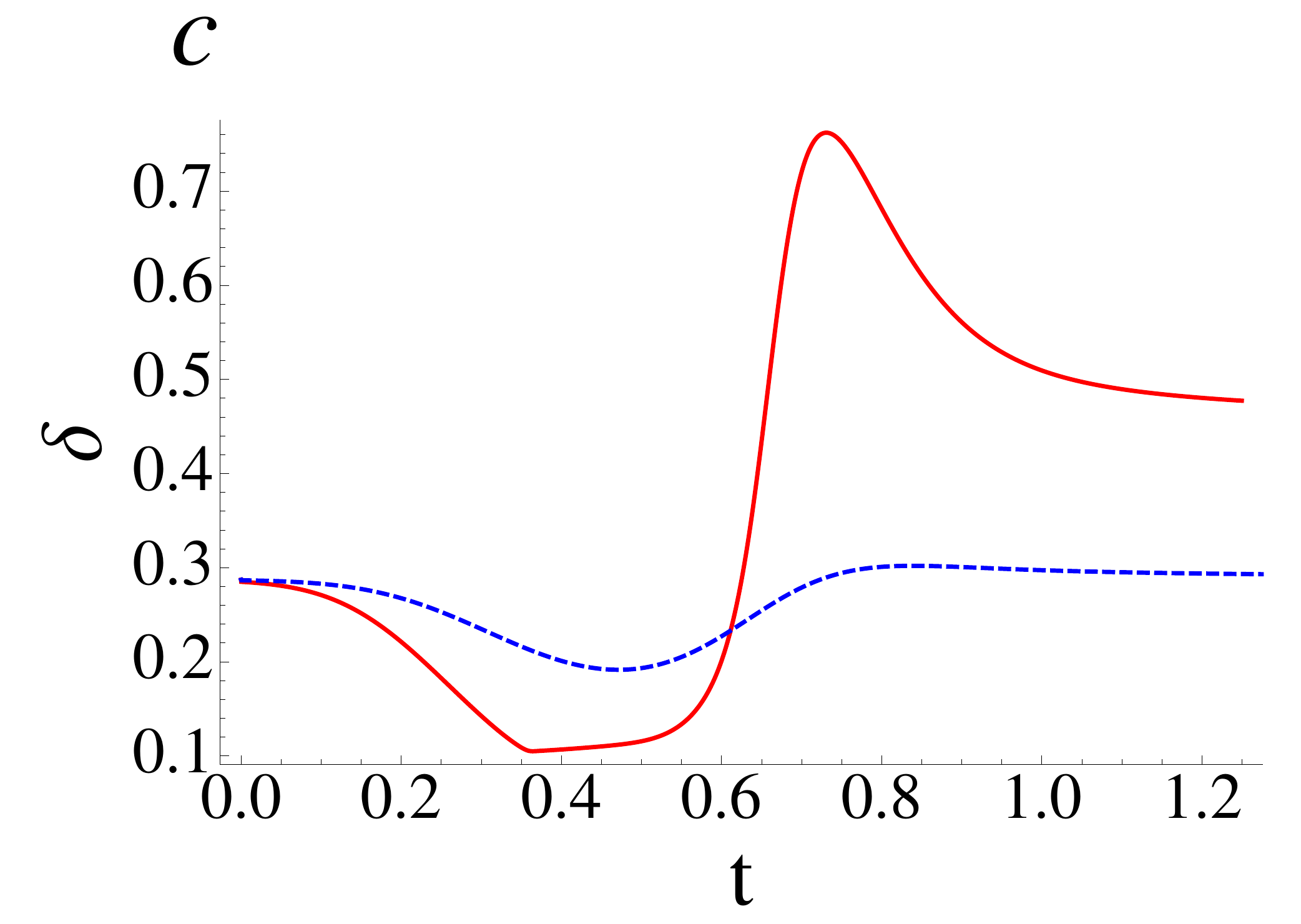}
	\includegraphics[width=7 cm]{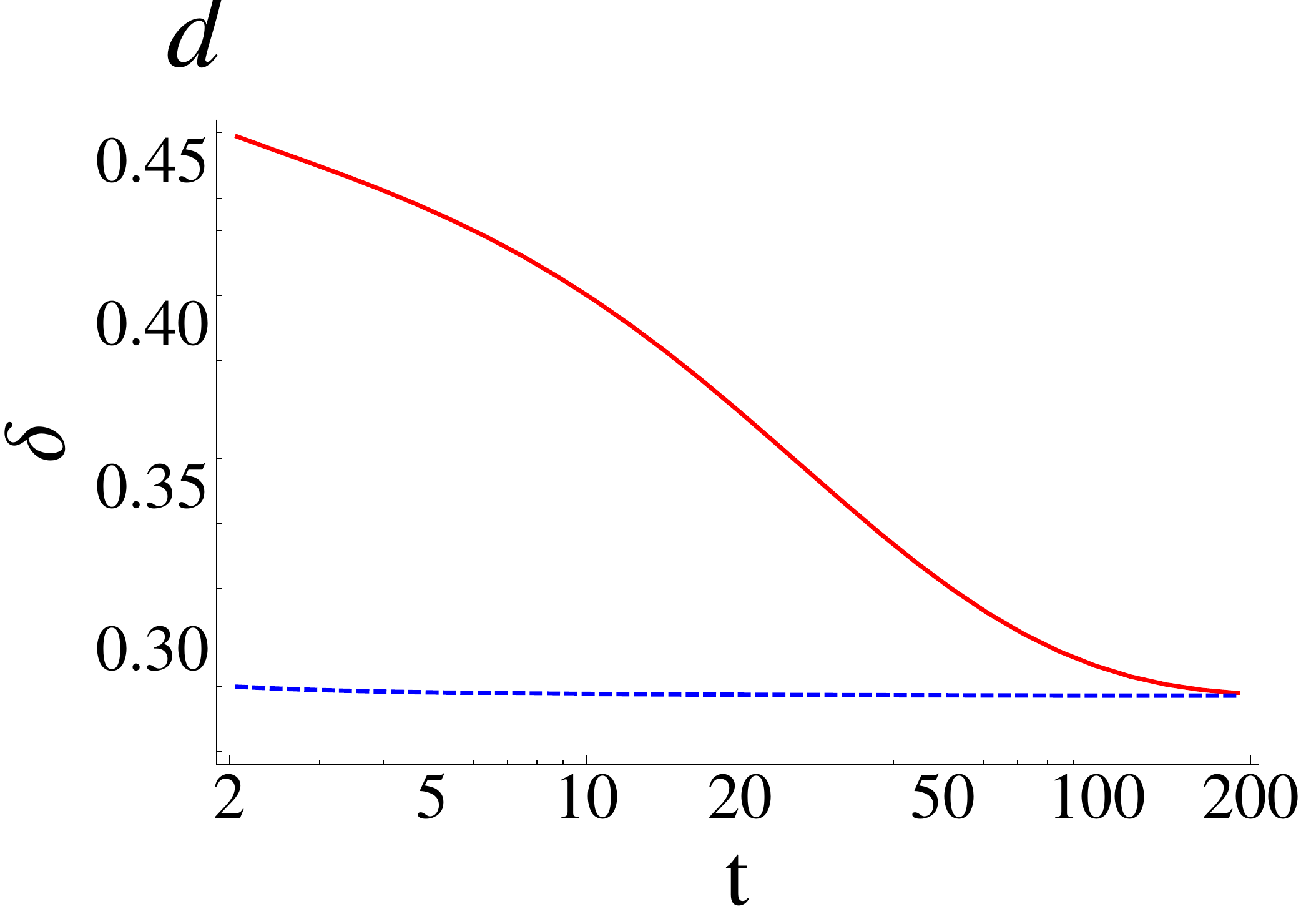}
	\caption{The internal force $f$ and the bond fraction $\nu$
largely determine the storage modulus $g'$ and the loss angle $\delta$:
normalized storage modulus $g'_{\rm n}$ ($a$) and loss angle $\delta$
($c$) during displacement pulses of large and small amplitudes, 2.2
(solid), 0.73 (dashed); the stronger deformation pulse causes a strong and
persistent overshoot of the loss angle, indicating {\em fluidization};
slow recovery after the pulse, ($b$, $d$); parameters as in figure
\ref{fig:force_displacement}, $\omega_0=3$.}
	\label{fig:modulus}
\end{center} \end{figure}

The substantial changes in the material properties accompanying bond
breaking can be exemplified by monitoring the linear elastic modulus
$g'(\omega_0)$ (measured at a fixed frequency $\omega_0$) in response to a
strain pulse (figure \ref{fig:modulus}a). Apart from the usual
\WLC/\GWLC-typical stress-stiffening below the threshold force $f_{\rm
y}$, the modulus is seen to drop \emph{below} the value it had before the
deformation pulse, where it apparently saturates (remember that the
deformation vanishes for $t>1$ and that the recovery takes roughly a time
$\tau_0e^{\eps-U}\gg 1$).  We thus observe what we call a ``passive'',
``physical'' remodelling of the bond network as opposed to the ``active'',
``biological'' remodelling of the cytoskeleton of living cells in response
to external stimuli. These  passive remodelling effects have recently been
observed for human airway smooth muscle cells \cite{TrepatDeng:2007}. (A
more quantitative discussion of the relation to the experimental
observations will be given elsewhere.) The fact that the deformation pulse
leads to a decrease in $g'$ and an increase in the loss angle $\delta$
(figure \ref{fig:modulus}c) suggests the notion of {\em fluidization}
\cite{TrepatDeng:2007}.

The passive remodelling described here is a {\em transient} phenomenon,
because, after cessation of the external perturbations, the bond fraction
will eventually recover its equilibrium value. This indicates that also
the change in the system properties is a transient effect, which is
demonstrated by the recovery from fluidization in figure~\ref{fig:modulus}
b \& d.  Thus, while the fluidization
resembles a plastic process on short time scales, on long times scales,
the phenomenology is more similar to pseudo- and superelasticity as
observed in shape-memory alloys \cite{DelaeyKrishnan:1974}.  

\begin{figure}[h] \begin{center}
	\includegraphics[width=14 cm]{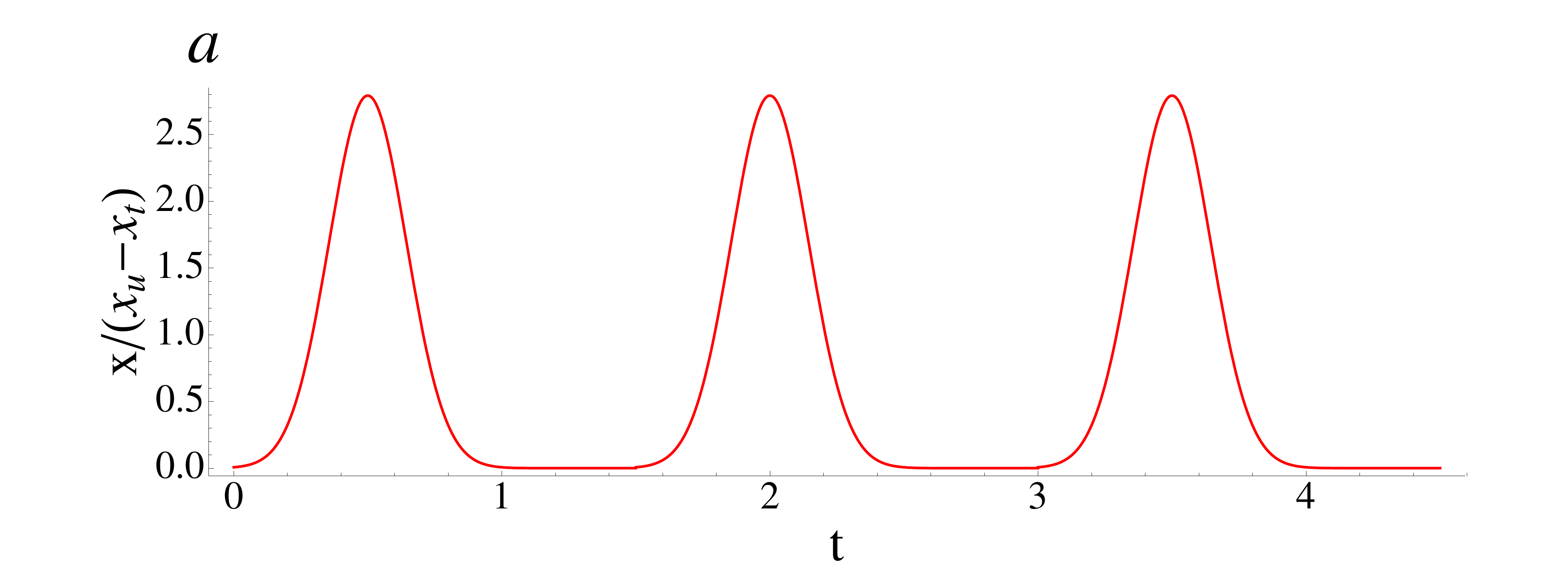}
	\includegraphics[width=7 cm]{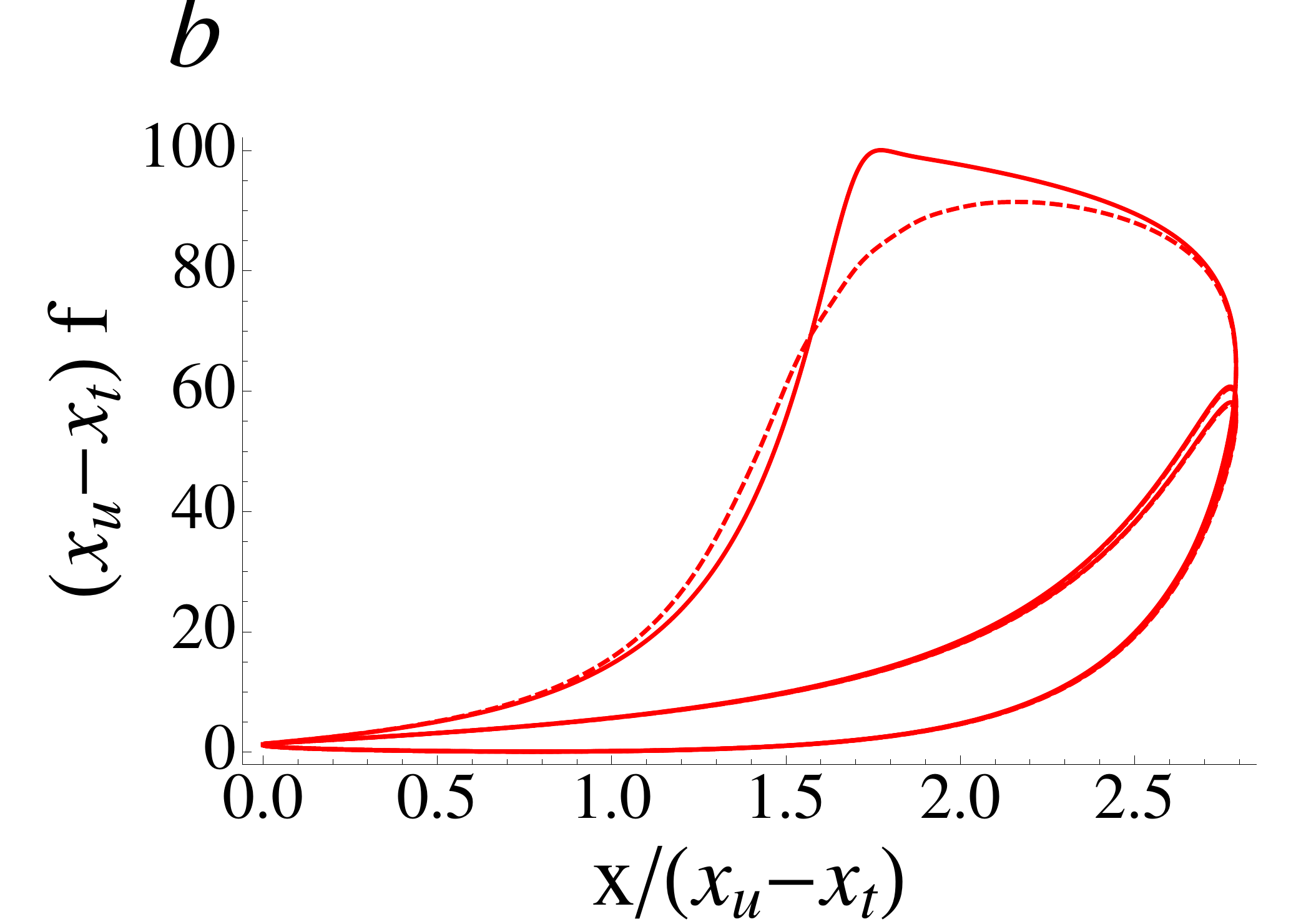}
	\includegraphics[width=7 cm]{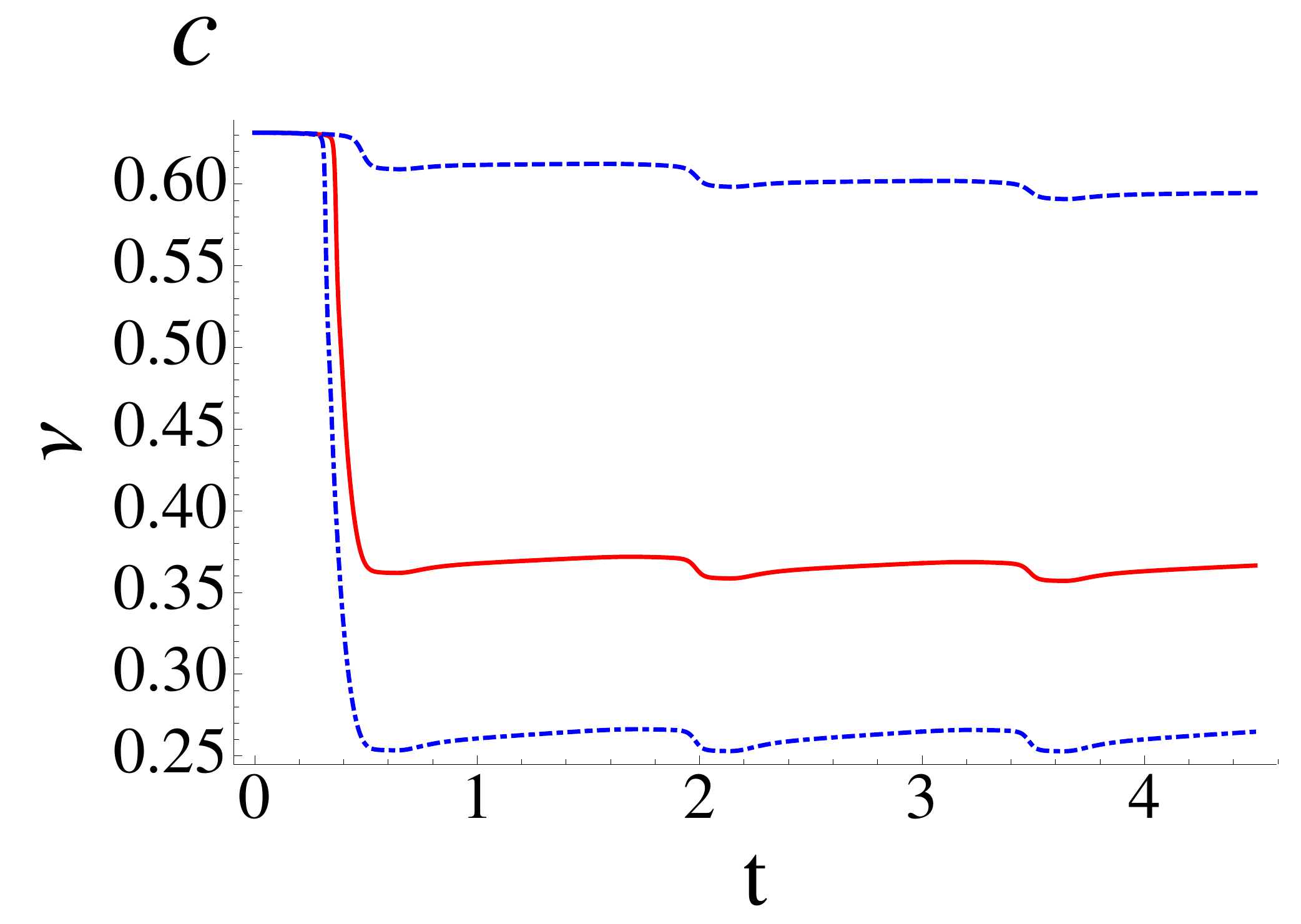}
	\caption{Protocol ($a$) and response ($b$, $c$) for three
          subsequent Gaussian strain pulses; force-displacement curves
          ($b$) for a single minimum interaction distance ($\Lambda_0=
          22.3$, solid) and for a distribution of $\Lambda_0$
          according to \cite{GlaserChakrabo:2009} with mean
          $\overline{\Lambda_0}= 22.3$ (dashed); the bond fraction
          $\nu$ ($c$) for $\Lambda_0=0.5 \,\overline{\Lambda_0}$
          (dash-dotted), $\overline{\Lambda_0}$ (solid), and $2
          \overline{\Lambda_0}$ (dashed); other parameters as in
          figure \ref{fig:force_displacement}. }
	\label{fig:stress_strain}
\end{center}\end{figure}

To demonstrate that the transient passive remodelling also
  affects the nonlinear material properties, we consider a series of
three pulses, next (figure \ref{fig:stress_strain}).  The force
response to such a protocol is depicted in figure
\ref{fig:stress_strain}b. The force-displacement curve for the second
stretching (second left branch of the solid curve) is less steep than
for the first stretching (very left branch of the solid curve) and the
yield force is lower. This indicates a cyclic softening, or
viscoelastic shake-down effect, closely related to the fluidization of
the network by strain. The strength of the shake-down depends on the
fraction of transiently broken bonds, and hence on the rate as well as
on the amplitude of the applied deformation (c.f.\ figure
\ref{fig:stress_strain}c, solid red curve). Upon repeated application
of deformation pulses the force-displacement curves settle on a
limit-curve that is essentially preconditioned by the initial
deformation pulse and the inelastic work it performed on the
sample. For more gentle protocols that only break a smaller fraction
of bonds, the initial fluidization would be not as pronounced and one
would obtain a gradual shake-down, which converges to a limit
  curve only after many cycles.

\subsection{Introducing multiple length scales}

\label{sec:intr_lengths}

So far, we have assumed that there is one well-defined
  characteristic length scale $\Lambda_0$ for the polymer
  interactions, which is on the order of the entanglement length $L_e$
  and interpreted as the minimum average contour distance between
  adjacent bonds of the test polymer with the background
network. This is clearly a mean-field assumption. Recent combined
experimental and theoretical studies have established that the local tube
diameter and entanglement length in pure semiflexible polymer
solutions actually exhibit a skewed leptocurtic distribution with
broad tails \cite{GlaserChakrabo:2009}.  To include this
information in the above analysis is not straightforward, since the
elastic interactions between regions of different entanglement length
are not known, \emph{a priori}.  For the sake of a first qualitative
estimate, the simplest procedure seems to be to average the above
results over a distribution of $\Lambda_0$,
corresponding to a parallel connection of independent entanglement
elements. Qualitatively, this renders the abrupt transition from
stiffening to softening somewhat smoother (c.f.\ figure
\ref{fig:stress_strain}b, dashed curve). Moreover, the initial
stiffening is slightly more pronounced. This is due to the
contributions of $\Lambda_0< \overline \Lambda_0$, which exhibit both
stronger stiffening and fluidization.  Nevertheless, all qualitative
features --- like stiffening and remodelling effects and a ``flow
state'' indicating fluidization or inelastic shakedown --- can still
be well discerned. This is consistent with the assumption that pure
semiflexible polymer solutions are at least qualitatively well
described by a single entanglement length
\cite{GittesMacKinto:1998,BauschKroy:2006,KroyGlaser:2007,Morse:2001,HinschWilhelm:2007}.

\section{Conclusions and outlook}

\label{sec:conclusions}

We have presented a theoretical framework for a polymer-based
description of the inelastic nonlinear mechanical behaviour of sticky
biopolymer networks. We represented the polymer network on a
mean-field level by a test polymer described by the
  phenomenologically highly successful \GWLC\ model. Beyond the
  equilibrium statistical fluctuations captured by the original model,
  we additionally allowed for a dynamically evolving thermodynamic
  state variable $\Lambda(t)$ characterizing the network of
  thermo-reversible sticky bonds, which is updated dynamically upon
  bond breaking according to an appropriate rate equation.  Within
our approximate treatment we could obtain a number of robust and
qualitatively interesting results, which are not sensitive to the
precise parameter values chosen, nor to the choice of the transverse
rather than the longitudinal susceptibility in (\ref{eq:sus_wlc}). For
sufficiently strong deformations, changes in the mean fraction of
closed bonds, reflected in $\Lambda(t)$, can influence material
properties to a degree at which the material behaves qualitatively
different from the usual viscoelastic paradigm. In particular, our
theory predicts a pronounced fluidization response, which develops
upon strong deformations on top of the intrinsic viscoelastic
stiffening response provided by the individual polymers. After
cessation of loading, the system slowly recovers its initial
state. These observations are in qualitative agreement with
recently published experimental results obtained for living cells
\cite{TrepatDeng:2007}, and a more quantitative comparison with
dedicated measurement results for cells and \emph{in vitro} biopolymer
networks will be the subject of future work. We found the yield force
for the onset of the fluidization to be sensitive to the deformation
rate.  Moreover, the fluidization response was shown to be
accompanied by a cyclic softening or shake-down effect.  Taking into
account the spatial heterogeneity of biopolymer solutions by a
distribution of entanglement lengths leads to a smoothing
of the force-displacement curves without affecting their qualitative
characteristics. One may still expect qualitatively new effects in
situations with unusually broad distributions of entanglement lengths,
such as for strongly heterogeneous (e.g.\ phase separating) networks.

A parameter that was found to be very important for cells
\cite{WangTolic-No:2002} but has not been discussed much in this
contribution is the {\em prestressing force} $f_0$ (see appendix
\ref{sec:technical}), which is present in adhering cells even in
the absence of external driving. Experimentally, it was established
that increasing prestress is correlated with a higher stiffness of
adhering cells \cite{WangTolic-No:2002} which in turn is correlated  with
a higher stiffness of the substrate \cite{DischerJanmey:2005}. When
naively treating the prestressing force just as an additive contribution
to the overall force, our model predicts the opposite: the additional
force breaks more bonds and the network becomes softer and more
fluid-like. To reconcile these apparently contradicting trends, one can appeal
to the notion of \emph{homeostasis}, which essentially amounts to
postulating that the cell actively adapts its structure such that a
certain set of thermodynamic state variables remain in a physiologically
sensible range \cite{WolffKroy:2009}.  This basically implies that the
cell will respond to stiff substrates or persistent external stresses by a
biological remodelling that corresponds to an increase of $\eps$ and
$U$, and possibly to a decrease of $\Lambda_0$. By adapting also the
internal prestress $f_0$, the cell may then avoid, apart from the
structural collapse, an equally undesirable loss of flexibility. We note,
however, that internal stresses are indeed observed to disrupt the
cytoskeleton, as implied by the simple physical theory presented here, if
they are not permanently balanced by a substrate
\cite{SalbreuxJoanny:2007}.

Even though the nonlinear effects presented in this contribution depend
crucially on the bond kinetics, it should not be overlooked that the
\GWLC\ is an equally indispensable ingredient of the complete theory.
First, it provides the constitutive relation that gives the force in
response to an infinitesimal displacement, which in turn governs the bond
kinetics. Secondly, the time-dependent linear susceptibility strongly
filters the dynamic remodelling of the bond network in practical
applications, in which one rarely has access to the microstate of the
underlying bond network. In summary, the extended model thus integrates
experimentally confirmed features of the \GWLC\ response such as slow
relaxation, power-law rheology, viscoelastic hysteresis, and shear
stiffening with a simple bond kinetics scheme, which allows less intuitive
complex nonlinear physical remodelling effects like fluidization and
inelastic shake-down to be addressed.

Finally, we want to point out that it is straightforward to extend the
above analysis in a natural way to account for irreversible plastic
deformations \cite{wolff-bullerjahn-kroy:tbp}. To this end, one has to
allow for the possibility that the transiently broken bonds reform
somewhere else than at their original sites (as always assumed in the
foregoing discussion) after strong deformations with finite residual
displacement.  This can be included by accommodating an additional
term acting as a ``slip rate'' in (\ref{eq:rate})
\cite{wolff-bullerjahn-kroy:tbp}.

\ack We acknowledge many fruitful discussions with J.~Glaser,
S.~Sturm, and K.~Hallatschek, and financial support from the Deutsche
Forschungsgemeinschaft (DFG) through FOR 877 and, within the German
excellence initiative, the Leipzig School of Natural Sciences ---
Building with Molecules and Nano-Objects.

\appendix

\section{Obtaining force-displacement curves and solving for $\nu$}

\label{sec:technical}

The deformation pulse protocol used in this paper is
\begin{equation}
	\Gamma(t)=\gamma_0 \exp\left(-\frac{(t-\tau)^2}{\sigma^2}\right).
\label{eq:strain_gauss}
\end{equation}
To obtain the full nonlinear response of the (extended) \GWLC\ to the
displacement protocol (\ref{eq:strain_gauss}), we make use of the
superposition principle: We know that the Fourier transform of
$g_{\mbox{\scriptsize \GWLC}}(\omega,\Lambda,f)/(i \omega)$ is the
linear response to a small displacement step.  After decomposing a finite
displacement into infinitesimal displacement steps and a partial
integration, we write
\begin{equation}
  f_{\mbox{\scriptsize \GWLC}}(t)= \int_{-\infty}^{\infty} \!\!\!dt'
\int_{-\infty}^{\infty}\frac{d\omega}{\sqrt{2 \pi}} \;
g_{\mbox{\scriptsize \GWLC}}\left(\omega,\Lambda(t),f_{\mbox{\scriptsize
\GWLC}}(t) \right)e^{-i \omega (t-t')} \Gamma(t').
\label{eq:force}
\end{equation}
Note that the right hand side of (\ref{eq:force}) depends on
$f_{\mbox{\scriptsize \GWLC}}(t)$, rendering it a highly nonlinear
implicit equation. Using (\ref{eq:strain_gauss}) and integrating by
parts, we obtain
\begin{equation}
		f_{\mbox{\scriptsize \GWLC}}(t)=\sqrt 2\gamma_0
\sigma \int_{0}^{\infty}\!\!\!\! d\omega\, e^{-i \omega
(t-\tau)}g_{\mbox{\scriptsize \GWLC}}(\omega,\Lambda(t),f_{\mbox{\scriptsize
\GWLC}}(t)) e^{-\frac{\sigma^2 \omega^2}{4}}.
\label{eq:force_gauss}
\end{equation}

The next step is to identify a connection between the force in
(\ref{eq:rate}) and $f_{\mbox{\scriptsize \GWLC}}$(t). A finite
prestressing force $f_0$ is introduced mainly for technical reasons,
namely to avoid the unphysical region of negative (i.e.\ compressive)
backbone stress, which would buckle the polymers. While the physics of
the prestressed network can essentially be mapped back to that of a
network without prestress by a renormalization of the parameters
$\eps$ and $U$, prestress may also be seen as an important feature:
indeed, the cytoskeleton of adhered cells is well known to be under
permanent prestress, and suspended cells seem prone to spontaneous
shape oscillations that could be indicative of a propensity of the
cells to set themselves under prestress \cite{WolffKroy:2009}.  In the
context and on the longer time scales of the biological remodelling of
the cytoskeleton it would therefore make sense to think of $f_0$ as a
dynamic force generated by molecular motors and polymerization
forces. For the following, however, we take the prestress to be
constant so that the force entering (\ref{eq:rate}) is
\begin{equation}
	f(t)=f_0+f_{\mbox{\scriptsize \GWLC}}(t).
\end{equation}
We now face the problem that (\ref{eq:force}) is an implicit equation
for $f_{\mbox{\scriptsize \GWLC}}(t)$, which depends on $\nu(t)$,
which in turn depends via $f(t)$ on $f_{\mbox{\scriptsize
    \GWLC}}(t)$.  We therefore use a two-step Euler scheme: As initial
values for $f$ and $\nu$, we choose the prestressing force,
$f(t=0)=f_0$, and the steady-state value under the prestressing force,
$\nu(t=0)=\nu(t\to\infty\vert f=f_0)$, respectively.  We then choose a
sufficiently small time step $\Delta t$ and apply the following
iterative rule:
\begin{eqnarray}
	f(k \Delta t)&\approx& f_0+f_{GWLC}\left(k \Delta t\vert f((k-1)
\Delta t,\nu((k-1) \Delta t)\right) \\
	\dot\nu(k  \Delta t)&\approx&\dot\nu(k  \Delta t\vert f(k \Delta
t)),
\end{eqnarray}
where $k \in \mathbb{N}_{+}$.
In the limit $\Delta t\to 0$, this procedure converges to the exact
solution.


\end{document}